\documentstyle[preprint,aps]{revtex}
\draft
\tighten
\input epsf
\font\blackboard=msbm10
\font\blackboards=msbm7
\font\blackboardss=msbm5
\newfam\black \textfont\black=\blackboard
\scriptfont\black=\blackboards
\scriptscriptfont\black=\blackboardss

\def\be{\begin{equation}}
\def\ee{\end{equation}}
\def\baray{\begin{eqnarray}}
\def\earay{\end{eqnarray}}

\def\N5{{1\over 2\times 5!}}

\begin{document}

\title{On the Degeneracy Inherent in Observational Determination
of the Dark Energy Equation of State}
\author{Ira Wasserman\footnote{ira@astro.cornell.edu}}
\address{Center for Radiophysics and Space Research\\
Cornell University\\
Ithaca, NY 14853}
\medskip
\date{\today}
\maketitle

\begin{abstract}
Using a specific model for the expansion rate of the Universe
as a function of scale factor, it is demonstrated that the
equation of state of the dark energy cannot be determined
uniquely from observations at redshifts $z\lesssim{\rm a~few}$
unless the fraction of the mass density of the Universe in
nonrelativistic particles, $\Omega_M$, somehow can be found
independently. A phenomenological model is employed to discuss
the utility of additional constraints from the formation of
large scale structure and the positions of CMB peaks in
breaking the degeneracy among models for the dark energy.

\end{abstract}
\section{Introduction}
\label{intro}

Although observations of acoustic peaks in the Cosmic
Microwave Background (CMB) fluctuation spectra
are sensitive to the curvature of spacetime, 
and appear to require some form of cosmological dark energy 
\cite{{2000Natur.404..955D},{2001PhRvD..63d2001L},{2001PhRvL..86.3475J},{DASI},{2001ApJ...561L...7S},{2002ApJ...564..559D}},
they may be less useful for discriminating 
among different equations of state for the dark energy.
Various studies (e.g. \cite{{1997PhRvD..56.4439T},{1997PhRvD..56.4625S}})
have concluded that the signatures of different types of
spatially smooth, evolving dark energy
on the CMB spectra are relatively undiscriminating, 
in part because CMB fluctuations were formed at high redshift,
when the dark energy was not a prominent constituent,
although spatial fluctuations in the dark energy could lift the
degeneracy considerably
\cite{1998PhRvL..80.1582C}.
By contrast, observations of sources at small to moderate redshifts
($z\lesssim{\rm a~few}$)
would probe epochs where the dark energy is prominent, leading to
suggestions that observations of Type Ia supernovae
\cite{{1998ApJ...509...74G},{2000ApJ...532..109P},{2000PhRvL..85.1162S},{1999PhRvD..60h1301H},{2001PhRvD..64l3527H},{2000PhRvD..62l1301C},{2001ApJ...550....1N}}
or of galaxy counts
\cite{{2001ApJ...563...28P},{2000ApJ...534L..11N},{2002ApJ...564..567N}}
could be used to determine the nature of the dark energy most
effectively.
However, the efficacy of these observational programs is controversial,
as there is considerable degeneracy among the predictions of
different dark energy models for, 
for example, the luminosity distance as a function of
redshift\cite{2001PhRvL..86....6M}.

One approach to the analysis of low redshift data would be to 
presume particular classes of models, perhaps parametrized by
a series expansion of $w(z)=p(z)/\rho(z)$, and then attempt to
constrain the model parameters by a likelihood or Bayesian method (e.g.
\cite{{1998ApJ...509...74G},{2000ApJ...532..109P},{2001ApJ...552..445W},{2001PhRvL..86....6M}}).
A second approach is to fit the data by a parametric 
representation of, for example, the luminosity distance as a 
function of redshift, and then analyze the results to constrain
the properties of the dark energy indirectly
(e.g. \cite{{2000PhRvL..85.1162S},{1999PhRvD..60h1301H},{2001PhRvD..64l3527H},{2000PhRvD..62l1301C},{2001ApJ...550....1N}}).
Although the first approach is preferable (because it generally
avoids difficulties associated with differentiating data, and
allows a simpler assessment of uncertainties in derived parameters, in
addition to outlining its implicit assumptions more clearly), 
the second approach is more useful for understanding whether
or not such observations can ever yield a unique solution
for the dark energy equation of state.

In this short note, the question considered is: Suppose one could
analyze data from $z\lesssim {\rm a~few}$ to
determine the expansion rate of the Universe as
a function of scale factor, $H(a)$, {\it exactly}. Could one then
determine the equation of state of the dark energy exactly from these
measurements alone? The answer, as we shall see, is {\it no}: a
{\it separate} determination of the fraction of the Universe in the
form of nonrelativistic particles is needed \cite{comment}.

\section{A Simple Illustrative Model}
\label{model}

The necessity of separately determining the density of
nonrelativistic particles in order to deduce the equation of
state of the dark energy
can be demonstrated most easily by constructing a specific
example. Suppose the data show that the Universe is expanding 
according to the simple relationship
\be
H^2=H_0^2\left(\Omega_1+{\Omega_2\over a^3}\right),
\label{expand}
\ee
where $a\equiv (1+z)^{-1}$.
One (tempting) interpretation of this result would be that the Universe
consists of two components, nonrelativistic matter contributing a
fraction $\Omega_2$ of the closure density, and a cosmological constant
contributing a fraction $\Omega_1=1-\Omega_2$. However, this
interpretation is not unique: there are models involving a scalar field
$\phi$ with nonconstant effective potentials $V(\phi)$ that can
lead to Eq. (\ref{expand}).

To see this, we can construct an explicit model with both a scalar field
and nonrelativistic matter, so that
\be
H^2={8\pi G\rho_\phi\over 3}+{H_0^2\Omega_M\over a^3}
\ee
with $\Omega_M\neq\Omega_2$ in general. The scalar field energy
density is therefore
\be
\rho_\phi=\rho_0\left(\Omega_1+{\Omega_2-\Omega_M\over a^3}\right),
\ee
where $\rho_0=3H_0^2/8\pi G$. Differentiate $\rho_\phi$ with respect to
$a$ to find
\be
{d\rho_\phi\over da}=-{3\rho_0(\Omega_2-\Omega_M)\over a^4}
=-{3(\rho_\phi+P_\phi)\over a}
\label{conserve}
\ee
where the second equality follows from conservation of energy
for the scalar field. 
Thus, we find simply that $P_\phi=-\rho_0\Omega_1$, and
%
\baray
\dot\phi^2&=&\rho_\phi+P_\phi={\rho_0(\Omega_2-\Omega_M)\over a^3}
\nonumber\\
V(\phi)&=&{\rho_\phi-P_\phi\over 2}=\rho_0\left(\Omega_1
+{(\Omega_2-\Omega_M)\over 2a^3}\right).
\label{phidentify}
\earay
Note that the first of Eqs. (\ref{phidentify}) requires that
$\Omega_2\geq\Omega_M$.
We can determine $V(\phi)$ by solving the first of Eqs.~(\ref{phidentify})
for $\phi(a)$, inverting to find $a(\phi)$, and substituting into
the second of Eqs.~(\ref{phidentify}). 
Combining Eqs. (\ref{expand}) and the first of Eqs. (\ref{phidentify})
implies
\be
a^2\left({d\phi\over da}\right)^2={\dot\phi^2\over H^2}
={3(\Omega_2-\Omega_M)\over
8\pi G\Omega_2\left(1+\Omega_1a^3/\Omega_2\right)}~.
\ee
Define
\be
x={\Omega_1a^3\over\Omega_2}\qquad{\rm and}\qquad
\phi_g^2={\Omega_2-\Omega_M\over 24\pi G\Omega_2}~;
\label{xdef}
\ee
then if $\psi\equiv\phi/\phi_g$
\be
x^2\left({d\psi\over dx}\right)^2={1\over 1+x}~,
\ee
which has the general solution
\be
\psi=\psi_\infty\pm 2\sinh^{-1}\left({1\over\sqrt{x}}\right)~,
\ee
where $\psi_\infty$ is a constant of integration. The sign 
in this solution can be chosen arbitrarily; chosing the
$-$ sign gives a solution in which $\psi$ increases with
increasing $a$. If we let $\psi=\psi_0$ today, then we find
\be
\psi=\psi_0+2\sinh^{-1}\sqrt{\Omega_2\over\Omega_1}
-2\sinh^{-1}\sqrt{\Omega_2\over\Omega_1a^3}~,
\label{psisol}
\ee
which we may invert to find
\be
{1\over a^3}=\left[\cosh\left({\hat\psi\over 2}\right)
-\sqrt{1\over\Omega_2}\sinh\left({\hat\psi\over 2}\right)
\right]^2~,
\ee
where $\hat\psi\equiv\psi-\psi_0$.
The second of Eqs. (\ref{phidentify}) then determines
the potential to be \cite{sen}
\be
V(\phi)=\rho_0\left\{\Omega_1+{1\over 2}(\Omega_2-\Omega_M)
\left[\cosh\left({\hat\psi\over 2}\right)
-\sqrt{1\over\Omega_2}\sinh\left({\hat\psi\over 2}\right) 
\right]^2\right\}~.
\label{conspirepot}
\ee
Note that $\hat\psi\to 2\coth^{-1}\sqrt{1/\Omega_2}$ and
$V(\phi)\to\rho_0\Omega_1$ asymptotically. Also, when
$\Omega_2=1-\Omega_1\to 1$, the potential becomes
exponential: $V(\phi)\to{1\over 2}\rho_0(1-\Omega_M)
e^{-\hat\psi}$.

The key question here revolves around prior assumptions. 
The ``simplest'' interpretation of Eq. (\ref{expand}) is
that the energy density of the Universe is the sum of 
contributions due to nonrelativistic matter and a 
cosmological constant. However natural this interpetation
may seem to be, it only follows if one assumes {\it either}
that the dark energy is in the form of a cosmological
constant, or that the mass density in nonrelativistic
form has closure parameter $\Omega_2$. If one
is somewhat more agnostic on either of these points,
then a range of possibile interpretations accounts for
the data equally well. The degeneracy is not just a
property of models for which $H^2(a)$ is given by Eq. 
(\ref{expand}). If the data were to favor a model of
the form
\be
H^2(a)=H_0^2\left[\Omega_1g(a)+{\Omega_2\over a^3}\right],
\ee
with $g(a)$ some function, and
$\Omega_1g(1)+\Omega_2=1$, then without additional prior
assumptions the most one can say for sure is that
$
\rho_\phi=\rho_0\left[\Omega_1g(a)+{(\Omega_2-\Omega_M)
/a^3}\right]$,
which leaves open the possibility of a range of different
solutions for $V(\phi)$.

This model illustrates that if one makes no prior assumptions
about the form of the quintessence potential $V(\phi)$, then
it is impossible to fit low redshift
data alone to find $V(\phi)$ uniquely unless
$\Omega_M$ is known independently -- in this example, a
range of models, parametrized by the value of $f\equiv\Omega_M/\Omega_2$,
will fit equally well. Constraints on the value of $f$ limit the
range of acceptable models. Although minimally $f\leq 1$, we know
that this model would be unable to account for the formation
of large scale structure if $f$ is too small. 
If we assume that only the nonrelativistic matter can clump
(i.e. that the dark energy and relativistic matter remain smooth),
then, in linear theory, the density contrast $D(a)$ obeys the
equation (e.g. \cite{1980lssu.book.....P})
\be
{d^2D\over da^2}+\left(3+{d\ln H\over d\ln a}\right)
{1\over a}{dD\over da}-{4\pi G\rho_M\over H^2a^2}D=0~;
\label{linear}
\ee
Eq. (\ref{linear}) has both growing and shrinking modes,
$D_+(a)$ and $D_-(a)$, respectively. If we include radiation
(which is a known component of the Universe, and subdominant
today, but important in determining linear theory growth
rates), then Eq. (\ref{expand}) must be altered to
\be
H^2(a)=H_0^2\left(\Omega_1+{\Omega_2\over a^3}+
{\Omega_{rad}\over a^4}\right),
\label{hrad}
\ee
where $\Omega_1+\Omega_2+\Omega_{rad}=1$ and,
for three low mass or massless neutrinos, 
$\Omega_{rad}h_{0.7}^2\approx 8.5\times 10^{-5}$ 
if $h_{0.7}=H_0/(70~{\rm km~s^{-1}~Mpc^{-1}})$.
The growing mode mainly amplifies perturbations between
$a_{eq}=\Omega_{rad}/\Omega_2$ and $a_1=(\Omega_2/
\Omega_1)^{1/3}$. During this phase, $H^2\propto a^{-3}$,
and $D_+(a)\propto a^{\sigma_+}$, where
\be
\sigma_+(f)={1\over 4}\left(\sqrt{1+24f}-1\right).
\ee
The overall linear theory growth factor is $\simeq (a_1/a_{eq})
^{\sigma_+(f)}$, so the ratio of the growth factor for
$f\neq 1$ to its value for $f=1$ is approximately
\be
\left({a_1\over a_{eq}}\right)^{{1\over 4}\left(\sqrt{1+24f}-5\right)}
\approx \left({a_1\over a_{eq}}\right)^{-3(1-f)/5}~,
\ee
where the approximation assumes $1-f\ll 1$. Thus, we
estimate that for the growth factor to be within a factor
of two of its value for $f=1$, we must have
$1-f\lesssim 5\ln 2/3\ln(a_1/a_{eq})\simeq 0.12$. 

The solid line in Fig. 1 shows the ratio of the growth factor for
$\Omega_M\leq\Omega_2$ to its value for $\Omega_2=\Omega_M$, 
that is
\be
R_{\delta\rho/\rho}\equiv 
{D_+(1; \Omega_M,\Omega_2)\over D_+(1;\Omega_2,\Omega_2)}
\ee
as a function of $\Omega_M$ for $\Omega_2=0.5$, where 
$D_+(a;\Omega_M,\Omega_2)$ is the linear theory growth factor
evaluated at scale factor $a$, given $\Omega_M$ and $\Omega_2$. These results were
found by solving Eq. (\ref{linear}) numerically from $a\ll a_{eq}$
to the present day, assuming identical initial perturbation amplitudes.
(Relativistic particles were included as a smooth component via
Eq. [\ref{hrad}].) The numerical results show that for  
$R_{\delta\rho/\rho}\geq 0.5$, then we must require $\Omega_M\gtrsim 0.44$, 
corresponding to $f\gtrsim 0.88$, approximately the same as the analytic
estimate of the previous paragraph. (The estimate is accurate to about
20\% for $\Omega_2\gtrsim 0.1$.) 

Linear theory also predicts
velocity fluctuations proportional to the peculiar velocity
factor $T_+(a;\Omega_M,\Omega_2)=d\ln D_+(a;\Omega_M,\Omega_2)/d\ln a$. 
The dashed line in Fig. 1 shows
\be
R_{\delta v}={T_+(1;\Omega_M,\Omega_2)\over T_+(1;\Omega_2,\Omega_2)}~,
\ee
the ratio 
of the peculiar velocity factor at 
$a=1$ for $\Omega_M\leq\Omega_2$ to its
value for $\Omega_M=\Omega_2$. Although there is some dependence of
$R_{\delta v}$ on $\Omega_M$, it is not as extreme as for
$R_{\delta\rho/\rho}$. 

Thus, primarily from the requirement that
linear density perturbations can grow substantially from small
values in the early Universe, we know that 
if Eq. (\ref{expand}) were truly exact, $f$ must be close to one.
Imposing limits on the range of plausible $\Omega_M$
from independent, physical and phenomenological arguments
amounts to using prior information to restrict
the possibilities to be compared with the data
from observations at $z\lesssim{\rm a~few}$. 
(The use of prior information from large scale structure and
CMB anisotropies has been discussed in e.g. \cite{{2001PhRvD..64l3527H},{1999MNRAS.310..842E},{1999PhRvL..83..670P},{2002PhRvD..65d1302B}}.)
This does not eliminate the degeneracy implicit in
Eq. (\ref{expand}) altogether, but does diminish its significance
by constraining the range of acceptable values of $f$ severely.
However, the fact remains that if one were to try to determine
the equation of state of the dark energy from observations that,
hypothetically, yield Eq. (\ref{expand}),
then one can never hope to find a unique result
from an analysis that ignores constraints from the development of large
scale structure (or any other considerations that yield 
independent information
about $\Omega_M$, such as peculiar velocities, CMB fluctuations and
weak lensing).
Moreover, even if such restrictions are imposed, unless $\Omega_M$ is
determined {\it precisely} by them, some degeneracy remains in 
the construction of $V(\phi)$ from observations at $z\lesssim
{\rm a~few}$.
%
\newcounter{figcount}
\setcounter{figcount}{1}
\begin{center}
\epsfxsize=5.5in
\epsfbox{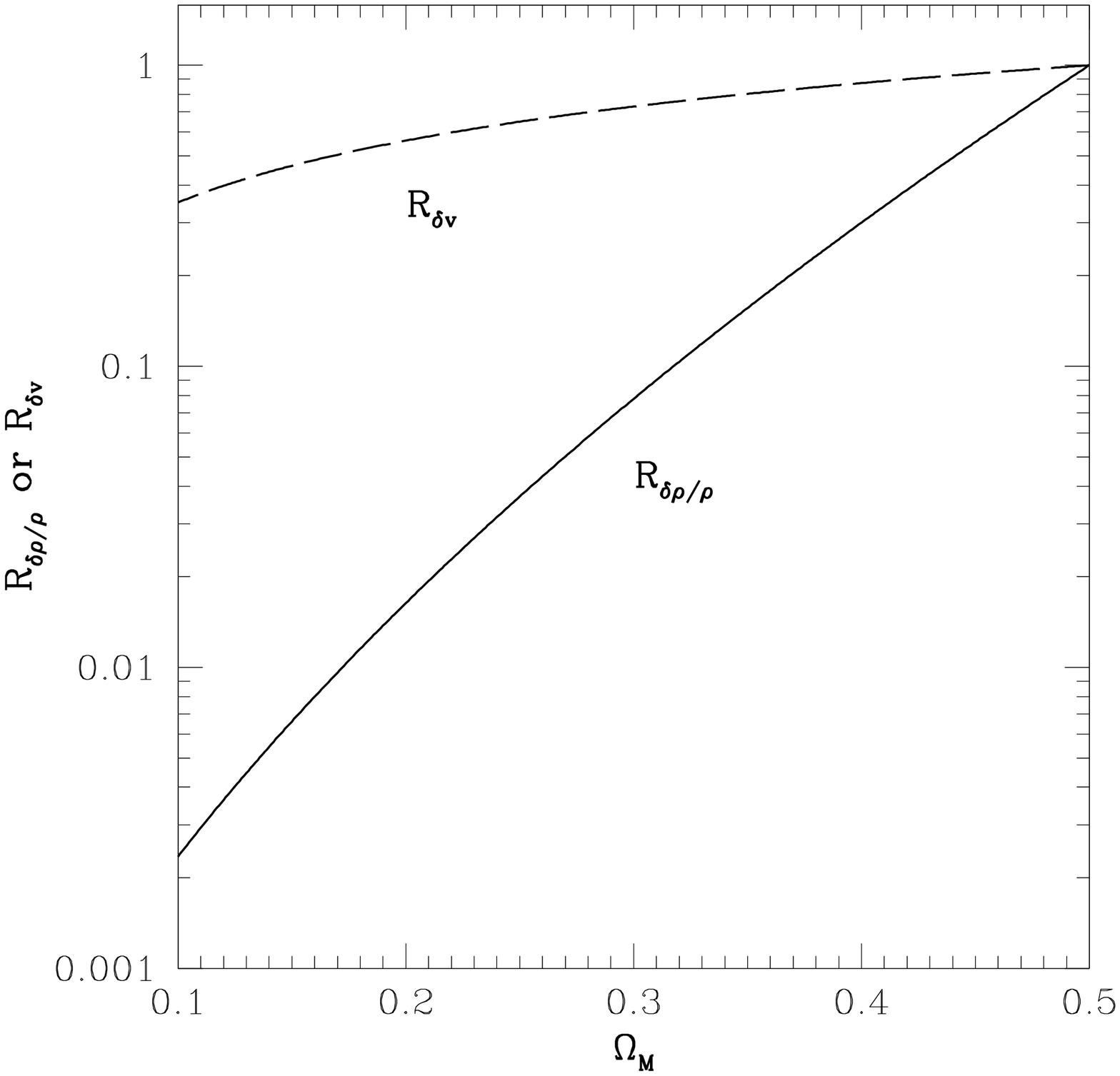}
\parbox{12cm}{\vspace{1cm}
FIG. \thefigcount\hspace{12pt}
Relative growth factor (solid) and velocity factor (dashed)
of linear perturbations
for models with $\Omega_M\leq\Omega_2=0.5$ that expand
according to Eq. (\ref{expand}).
\vspace{12pt}
}
\end{center}

\section{Can $\Omega_M$ and $\Omega_2$ Differ Substantially?}
\label{diff}

If Eq. (\ref{expand}) with $\Omega_2$
substantially different from
$\Omega_M$ is an excellent approximation at low to moderate
redshifts, it can {\it never} be exact. In this case, there must be a
transition in $H^2(a)$, so that the component of the density
that scales like $a^{-3}$ is predominantly due to nonrelativistic
particles for a period sufficient to grow large scale structure.
The implicit deviations from
Eq. (\ref{expand}) would contain information on $\Omega_M$ as
well as $\Omega_2$, and probably additional parameters as well.
In this situation, it might be possible to
ascertain the equation of state of the dark energy from measurements
at low to moderate redshifts alone, provided that the imprint of $\Omega_M$
and the additional parameters in $H^2(a)$ can be discerned from
these observations. The question is, if Eq. (\ref{expand}) is an accurate
(but inexact) fit at $z\lesssim{\rm a~few}$, and $\Omega_2$ differs
substantially from $\Omega_M$, how well could we determine $\Omega_M$,
and hence break the degeneracy, observationally?

To make these issues more concrete, suppose in reality Eq. (\ref{expand})
is actually the large $a$ limit of a more general, exact relationship, say
\be
H^2(a)=H_>^2(a)F\left({a\over a_t}\right)+H_<^2(a)
\left[1-F\left({a\over a_t}\right)\right],
\label{newexpand}
\ee
where $F(q)\to 0$ as $q\to 0$ and $F(q)\to 1$ as $q\to\infty$,
with $a_t<1$ a transition value of the scale factor, $H^2_>(a)$ given
by Eq. (\ref{expand}), and, for example,
\be
H^2_<(a)=H_0^2\left(\Omega_<+{\Omega_M\over a^3}\right)~;
\ee
since the model is flat, 
\be
1=(\Omega_1+\Omega_2)F_0+
(\Omega_<+\Omega_M)(1-F_0),
\label{flatcond}
\ee
where $F_0\equiv F(1/a_t)$. Such
a model could be consistent with the growth of large scale structure,
but might still be hard to distinguish from Eq. (\ref{expand}) based on
observations at low to moderate redshift, depending on the value of
$a_t$, and the form of $F(a/a_t)$. Eq. (\ref{newexpand}), although
admittedly {\it ad hoc}, is useful for studying the extent to which
observations of various sorts could discern that $\Omega_2\neq\Omega_M$,
and pin down the difference sufficiently to break the degeneracy discussed
above.

For a flat cosmology, Eq. (\ref{newexpand}) implies that the radial
coordinate for a source at $a=(1+z)^{-1}$ is
\baray
r_S(a;\Omega_M,\Omega_2,F)&=&\eta(1)-\eta(a)=
\int_a^1{db\over b^2H(b)}\nonumber\\
&=&H_0^{-1}\int_a^1
{db\over\sqrt{b\left\{\Omega_eb^3+1-\Omega_e
+[(\Omega_<-\Omega_1)b^3+\Omega_M-\Omega_2][F_0-F(b/a_t)]\right\}}}
\label{sourcecoord}
\earay
where $\Omega_e\equiv\Omega_1F_0+\Omega_<(1-F_0)$, and $d\eta\equiv dt/a
=da/a^2H(a)$. (Here, the notation $r_S(a;\Omega_M,\Omega_2,F)$
means the radial coordinate as a function of $a$ given $\Omega_M$ and
$\Omega_2\geq\Omega_M$, as well as a set of parameters required
to define $F(a/a_t)$.)
If $F(a/a_t)$ does not vary much over the range of $a$ covered by the
observations, then the data would only determine $\Omega_e$ accurately,
and would yield little useful
information on $\Omega_2$ or $\Omega_M$: only the combination
$\Omega_2F_0+\Omega_M(1-F_0)=1-\Omega_e$ could be deduced. This situation
is possible provided that $a_t$ is fairly small, below the range covered
by observations, and $F(a/a_t)$ varies relatively
slowly as a function of $a/a_t$ for large values of $a/a_t$. In this case,
we cannot expect to eliminate the degeneracy among models directly from
observations of low to moderate redshift sources alone. Constraints
from other sorts of observations that probe $\Omega_M$ directly
would be needed to deduce the quintessence model as accurately as possible.

More quantitative statements only can be made in terms of specific
$F(a/a_t)$, and require some exploration of parameters. Let us examine
one choice: $\Omega_1+\Omega_2=1=\Omega_<+\Omega_M$, in which case
Eq. (\ref{flatcond}) is satisfied exactly, and Eq. (\ref{sourcecoord})
becomes
\be
r_S(a;\Omega_2,\Omega_M,F)=H_0^{-1}\int_a^1{db\over\sqrt{b\left\{\Omega_2+(1-\Omega_2)b^3
-(\Omega_2-\Omega_M)(1-b^3)\left[1-F(b/a_t)\right]\right\}}}~.
\label{simplerad}
\ee
%
\setcounter{figcount}{2}
\begin{center}
\epsfxsize=5.5in
\epsfbox{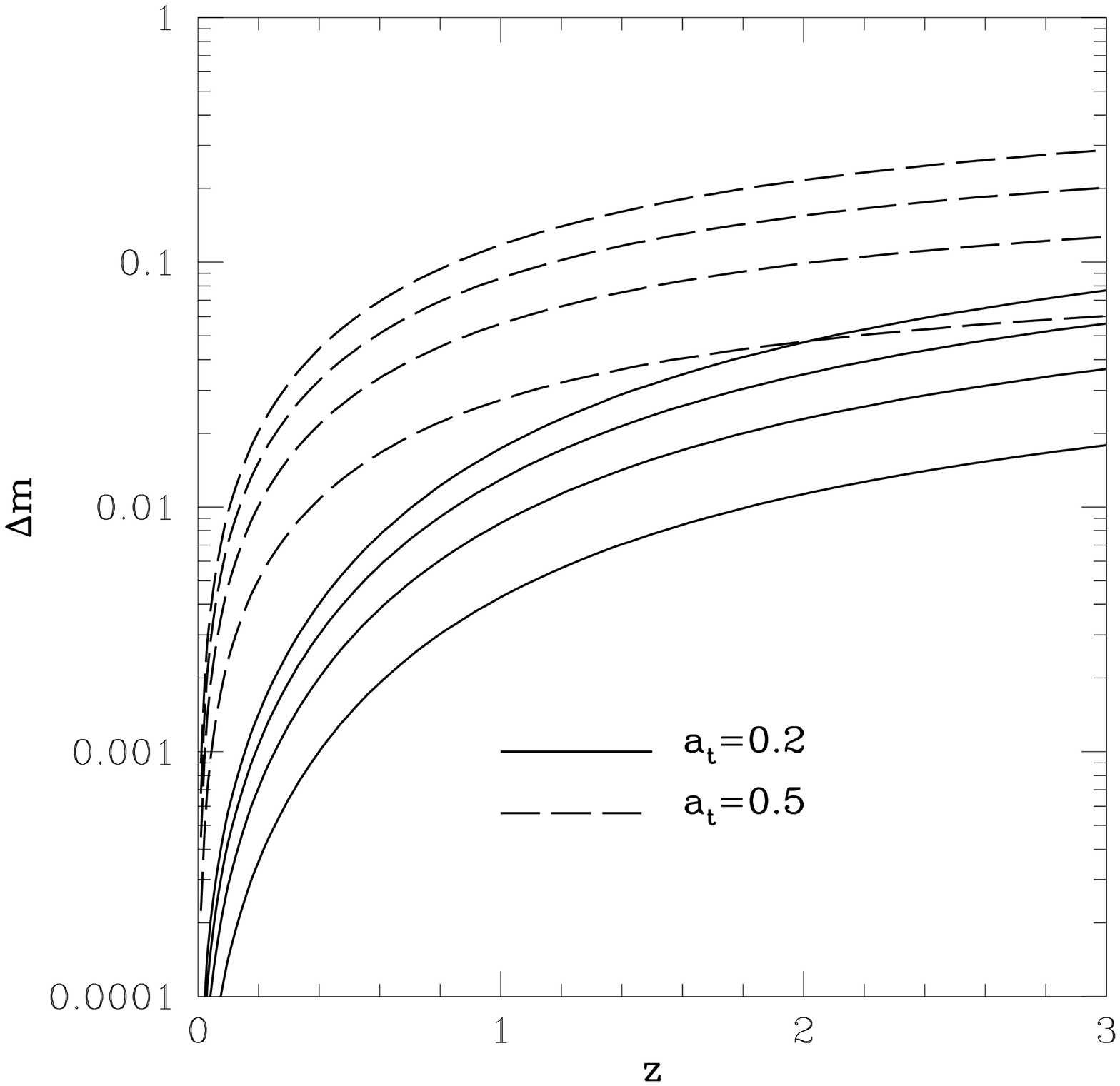}
\parbox{12cm}{\vspace{1cm}
FIG. \thefigcount\hspace{12pt}
Magnitude differences among models with $\Omega_2=0.5$ for
$a_t=0.2$ (solid) and $a_t=0.5$ (dashed) as functions of
$z$ for, from the top for each $a_t$, $\Omega_M=0.1, 0.2, 0.3$ and
$0.4$.
\vspace{12pt}
}
\end{center}

Eq. (\ref{simplerad}) may be used to compute magnitude differences
among models with given $\Omega_2$, $\Omega_M$ and $F(a/a_t)$; for
illustrative purposes, let us choose $1-F(a/a_t)=e^{-a/a_t}$.
Fig. ~2 depicts the magnitude difference
\be
\Delta m(z;\Omega_2,\Omega_M,a_t)=5.0\log_{10}\left[{r_S((1+z)^{-1};\Omega_M,\Omega_2,a_t)
\over r_S((1+z)^{-1};\Omega_2,\Omega_2,a_t)}\right]
\ee
relative to the model with
$\Omega_M=\Omega_2=0.5$
as a function of $z$ assuming $a_t=0.2$ (solid) and 
$a_t=0.5$ (dashed) for (from top to bottom in each case)
$\Omega_M=0.1, 0.2, 0.3$ and $0.4$.
Not surprisingly, the magnitude differences are smaller
for the larger value of $a_t$, and decrease as $\Omega_M\to\Omega_2$.
Nevertheless, the {\it largest} magnitude difference shown for 
$z\leq 3$ is 0.29 mag, for $(z,a_t,\Omega_M)=(3,0.5,0.1)$; for
$(z,a_t,\Omega_M)=(3,0.2,0.1)$, the difference is 0.077 mag. 

%
\setcounter{figcount}{3}
\begin{center}
\epsfxsize=5.5in 
\epsfbox{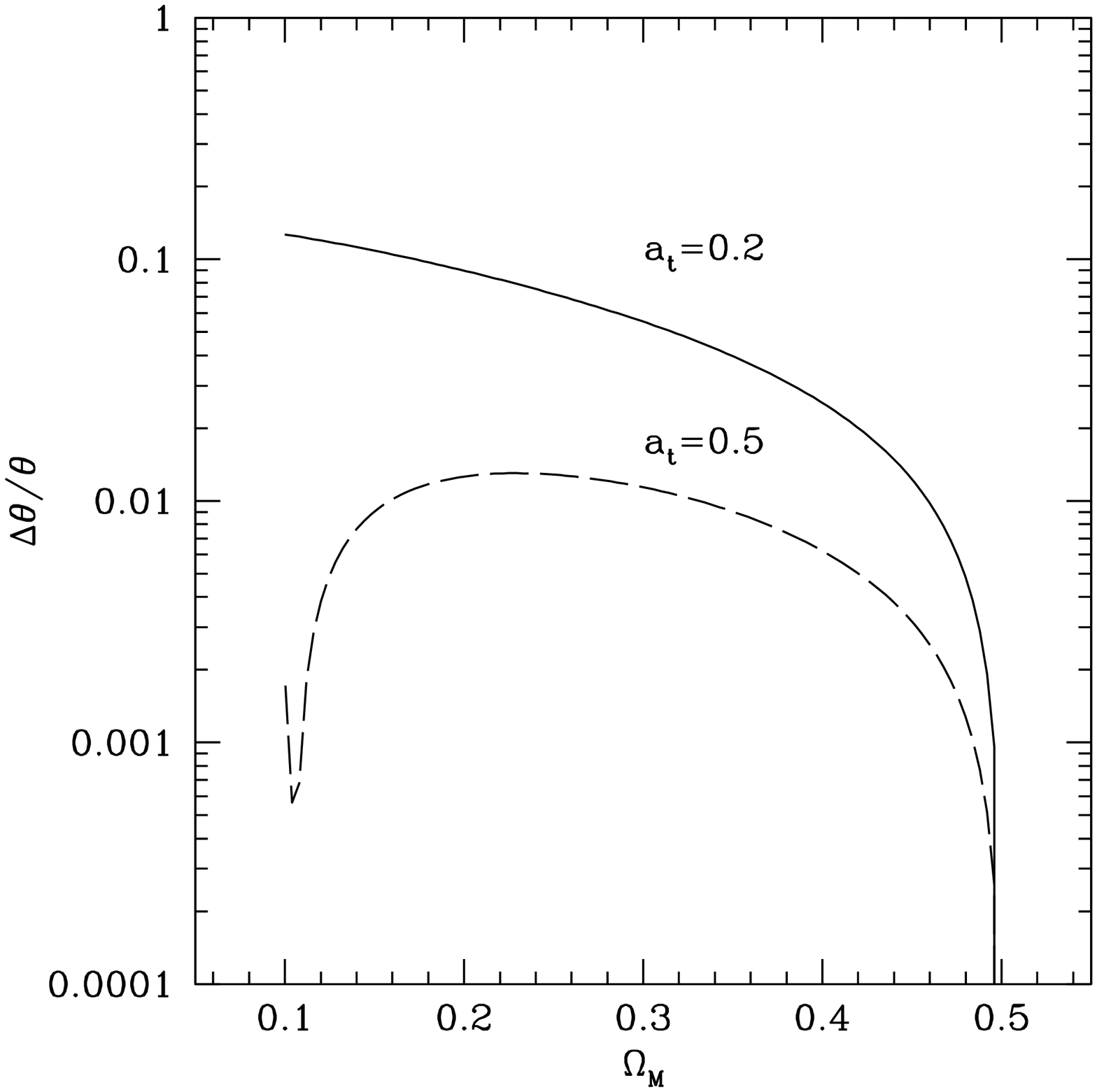}
\parbox{12cm}{\vspace{1cm}
FIG. \thefigcount\hspace{12pt}
Fractional deviations of the angular size of the 
(sonic) horizon at $a=10^{-3}$ from the $\Omega_M=
\Omega_2=0.5$ model as functions of $\Omega_M\leq
\Omega_2$. The solid line is for $a_t=0.2$, and
the dashed line for $a_t=0.5$. For computing the angular sizes,
$H_0=70~{\rm km~s^{-1}~Mpc^{-1}}$ was adopted.
\vspace{12pt}
}
\end{center}

Fig. ~3 shows results for
\be
{\Delta\theta\over\theta}
(\Omega_M;\Omega_2,a_t)=\left\vert{\eta(10^{-3};\Omega_M,\Omega_2,a_t)
r_S(10^{-3};\Omega_2,\Omega_2,a_t)
\over r_S(10^{-3};\Omega_M,\Omega_2,a_t)\eta(10^{-3};\Omega_2,\Omega_2,a_t)}-1\right\vert,
\label{rsmodel}
\ee
the fractional difference between the angular
size of the (sonic) horizon at $a=10^{-3}$ as a function of $\Omega_M
\leq\Omega_2$ and its value for $\Omega_M=\Omega_2$ for the
two different values of $a_t$. (Deviations of the sound speed from
$1/\sqrt{3}$ have not been included.) Larger fractional deviations arise
for smaller $\Omega_M$ and $a_t$. The largest deviation shown in
the two figures is 12.7\% for $(\Omega_M,a_t)=(0.1,0.2)$; the largest
fractional deviation for $a_t=0.5$ is about 1.3\%, and occurs near
$\Omega_M=0.225$. 

Although this is only one example, based on a particular ({\it ad hoc})
choice of $F(a/a_t)$,
it illustrates that while data at both low to 
moderate $z$ as well as CMB observations can yield information that
lifts the degeneracy between $\Omega_2$ and $\Omega_M$, considerable
precision may be needed to determine the equation of state of the
dark energy accurately. Note that no attempt has been made to fit simulated
data based on Eq. (\ref{rsmodel}) to figure out how well $H(a)$
could be determined in practice; the comparisons made in Figs. 2 and 3
are based on exact functions. The observational challenge of 
distinguishing among models at these levels could be even greater
than these relatively small differences would indicate.
On the other hand, Figs. 2 and 3 were constructed based on a particular
({\it ad hoc}) model for $H^2(a)$ designed for a phenomenological
study of how easily information on $\Omega_M$ could be gleaned
from luminosity distances and CMB acoustic peaks. The task could
be easier or harder in cosmologies based on particular physical
theories for the dark energy, depending on the details of the model.

\section{Discussion}
\label{discuss}

To conclude, the point of this paper is {\it not} to suggest that it
is absolutely inconceivable that the equation of state of the dark
energy could be measured, but rather to show, via a specific
example, how measurements at $z\lesssim 3$ 
admit a continuum of interpetations in terms of evolving dark energy
fields {\it unless} $\Omega_M$ can be determined separately somehow.
The specific example, Eq. (\ref{expand}),
can be intepreted ``naturally'' in terms of a
Universe containing a mixture of nonrelativistic particle
dark matter plus a cosmological constant, but can
be interpreted equally well in terms of a quintessence model
with a range of potentials, Eq. (\ref{conspirepot}).
This specific example sheds some light on why simulated analyses
of low to medium redshift data appear to allow degenerate
interpretations: it is always possible that the component of
the total energy density of the Universe that evolves like
$a^{-3}$ is only partially due to nonrelativistic particles,
with the rest arising from quintessence. Unique interpretations
can only be obtained 
in terms of specific models for the
quintessence that forbid such a conspiracy {\it a priori}, by either
specifying the form of $V(\phi)$ or fixing the value
of $\Omega_M$.

Realistically, the extent to which the $\rho\propto a^{-3}$
constituent must be due to nonrelativistic particles is
constrained by the requirement that large scale
structure formation evolve ``normally,'' that is, unimpeded
by the existence of a component with density proportional
to $a^{-3}$ that is incapable of clustering.
If Eq. (\ref{expand}) were truly exact, then $\Omega_2$
would have to be close to $\Omega_M$ for large scale structure
to grow, but even in this case, a limited degeneracy remains
in the determination of $V(\phi)$. If $\Omega_2$ and $\Omega_M$
differ substantially, then Eq. (\ref{expand}) cannot be exact,
raising the question of how well one could discern the two
parameters separately from observations. To examine
this issue, a modified (phenomenological)
expansion law, Eq. (\ref{newexpand}), that encodes information
on both $\Omega_2$ and $\Omega_M$
was introduced. In the context of a particular model that is
consistent with large scale structure formation, but reduces
to Eq. (\ref{expand}) with $\Omega_2\geq\Omega_M$ 
at large enough $a$, we have seen
that information about $\Omega_M$ could be 
gleaned both from measurements of the positions of CMB
peaks and from luminosity distance determinations. However,
the deviations among models with various $\Omega_M\leq\Omega_2$
can be quite small, which would still pose a substantial
challenge for programs that aim to determine the equation
of state of the dark energy. 

Of course, in the context
of {\it specific} models for the quintessence field,
embodied in particular forms for $V(\phi)$,
the analyses may not encounter pronounced
degeneracies. At present, though, there is little
compelling reason to assume any particular $V(\phi)$ 
{\it a priori}. When analyzing data for a particular form
of $V(\phi)$, one is primarily engaged in estimating
the parameters of the model, as well as $\Omega_M$. Thus,
one might be able to find the best fit model of a particular
type (e.g. $V(\phi)={\rm constant}$), without being able
to tell if the underlying model would be favored by the
data if other possibilities were admitted. If the allowed
ranges of parameters for a particular quintessence model
do not shrink with the accumulation of data, one can
conclude that the model is inadequate with confidence. 
However, there is no guarantee that a given model
is correct even if the data seem to converge on a
unique set of parameters,
as Eqs. (\ref{expand}) and (\ref{conspirepot}) illustrate.

\acknowledgments
I thank Eanna Flanagan, Daniel Holz, and Paul Steinhardt for helpful
comments on this paper.

\end{document}